\documentclass[onecolumn,12pt,nofootinbib]{revtex4}
\usepackage{graphicx}
\usepackage{dcolumn}
\usepackage{amsmath}
\usepackage{amssymb}
\usepackage{subfigure,amsmath,verbatim,moreverb,bm}

\newcommand{\dd}{\mathrm{d}}
\newcommand{\rv}{\mathbf{r}}

\newcommand{\xvu}{\underline{\mathbf{x}}}
\newcommand{\diamatA}{|\Psi^A_0|^2}
\newcommand{\diamatB}{|\Psi^B_0|^2}
\newcommand{\method}{\mathrm{FDM}}
\begin{document}
\title{London dispersion forces without density distortion: a path to first principles inclusion in density functional theory}
\author{Derk P. Kooi}
\affiliation{Department of Chemistry \& Pharmaceutical Sciences and Amsterdam Institute of Molecular and Life Sciences (AIMMS), Faculty of Science, Vrije Universiteit, De Boelelaan 1083, 1081HV Amsterdam, The Netherlands.}
\email{d.p.kooi@vu.nl}
\author{Paola Gori-Giorgi}
\affiliation{Department of Chemistry \& Pharmaceutical Sciences and Amsterdam Institute of Molecular and Life Sciences (AIMMS), Faculty of Science, Vrije Universiteit, De Boelelaan 1083, 1081HV Amsterdam, The Netherlands.}

\date{\today}

\begin{abstract}
We analyse a path to construct density functionals for the dispersion interaction energy from an expression in terms of the ground state densities and exchange-correlation holes of the isolated fragments. The expression is based on a constrained search formalism for a supramolecular wavefunction that is forced to leave the diagonal of the many-body density matrix of each fragment unchanged, and is exact for the interaction between one-electron densities. We discuss several aspects: the needed features a density functional approximation for the exchange-correlation holes of the monomers should have, the optimal choice of the one-electron basis needed (named ``dispersals''), and the functional derivative with respect to monomer density variations.
\end{abstract}
\maketitle


\section{Introduction}
The London Dispersion forces are crucial in physics, chemistry and biology; one of their most wonderous properties is that they make ``everything\footnote{Here it should be noted that this only holds for neutral and non-dipolar systems. However, for dipolar systems there is always an orientation such that the force between the two systems is attractive.} stick to everything".\cite{LieThi-PRA-86} Density Functional Theory (DFT) approximations have difficulties in describing dispersion interactions, which stem from the inherent non-locality of the phenomenon. Attempts at building non-local functionals to capture dispersion have been somewhat successful \cite{LanDioRyd-IJQC-04,BerCooSchThoHylLun-RPP-15,StoVooTka-CSR-19} especially in calculations on solids,\cite{Bjo-PRB-12,BerCooSchThoHylLun-RPP-15} but they are typically outcompeted in practical calculations on molecules by empirical corrections based on pairwise atomic dispersion coefficients derived from free atoms\cite{BurMaySum-JCP-11,CalBanGri-JCP-17} or on free atomic polarizabilities combined with dipole moments of the exchange hole.\cite{BecJoh-JCP-07} However, these empirical corrections are only able to take into account effects of the chemical environment of the atom in a limited way and do not model anisotropy in the dispersion coefficients. For very recent reviews on the topic, including different angles and perspectives, see, for example, refs~\citenum{DobGou-JPCM-12,GriHanBra-CR-16,StoVooTka-CSR-19}.\\
Within a DFT framework, a microscopic model of the dispersion mechanism based on ground-state properties only, is highly desirable, in order to provide a basis to build approximate correlation functionals able to capture dispersion. The exchange-hole dipole moment (XDM) model of Becke and Johnson \cite{BecJoh-JCP-07} was a step in this direction, which however, still needs the atomic polarizabilities as input. A good real-space modeling of the inherent mechanism behind dispersion is also important to address strongly correlated systems. For example, dispersion between localised $d$ or $f$ electrons can play an important role also at equilibrium geometries. \\
It is the purpose of this paper to provide and analyze a microscopic model of dispersion based on the ground-state densities and exchange-correlation holes of the fragments only. The framework is based on our recent work\cite{KooGor-JPCL-19} on a class of variational wavefunctions that capture the long-range interaction between two systems without changing the diagonal of the density matrix of each monomer. The advantage is that dispersion becomes in this way a simple competition between kinetic energy and monomer-monomer interaction, as all the remaining potential energy terms inside each monomer are not allowed to change and cancel out in the interaction energy. The formalism is thus analogous to the Levy-constrained search for the universal ground-state energy density functional,\cite{Lev-PNAS-79} as we will explain in sec~\ref{sec:theory}.\\
Although this variational wavefunction is certainly not exact by construction (see also the discussion in sec~\ref{sec:thm}), it can yield accurate or even exact results for the dispersion coefficients when combined with accurate pair densities of the monomers.\cite{KooGor-JPCL-19,KooWecGor-arxiv-20}
While in our previous work we mainly took a purely wavefunction perspective, in this article we focus primarily on the theoretical aspects of the use of our framework to build DFT approximations for dispersion, also discussing the (possible) violations of the Hellman-Feynman, Hohenberg-Kohn and virial theorems. We also report the functional derivative and discuss which features of the monomer xc-hole approximations are really needed. Finally, we analyze the one-body functions needed to expand our wavefunction, which we have named ``dispersals''. We look into the vector space of these dispersals and discuss equations for an optimal finite set of dispersals. We describe an approach to generate atomic dispersals, which can be combined into molecular dispersals to limit the computational cost of the procedure in future work.

\section{Theory}\label{sec:theory}
We consider two neutral systems (atoms, molecules) $A$ and $B$ separated by a large distance $R$, having isolated ground-state wavefunctions $\Psi_0^A(\xvu_A)$ and $\Psi_0^B(\xvu_B)$, with $\xvu_{A/B}$ denoting the set of all their electronic spin-spatial coordinates. 
We define the ``fixed diagonal matrices'' (FDM) dispersion energy $E^\method_{\rm disp}$ between $A$ and $B$ via the following constrained minimisation problem
\begin{equation}\label{eq:EdispLevy}
	E^\method_{\rm disp}(R)=\min_{\Psi_R\to \diamatA,\diamatB}\langle\Psi_R|\hat{T}+\hat{V}_{ee}^{AB}|\Psi_R\rangle-\langle \Psi_0^A\Psi_0^B|\hat{T}+\hat{V}_{ee}^{AB}|\Psi_0^A\Psi_0^B \rangle,
\end{equation}
where $\hat{T}$ is the usual kinetic energy operator acting on the full set of variables $\xvu_A,\xvu_B$, and
\begin{equation}\label{eq:VeeAB}
	\hat{V}_{ee}^{AB}=\sum_{i\in A,j\in B}\frac{1}{|\rv_i-\rv_j|}.
\end{equation}
The notation ${\Psi_R\to \diamatA,\diamatB}$ means that the wavefunction $\Psi_R(\xvu_A,\xvu_B)$ is forced to leave the diagonal of the many-body density matrix of each fragment unchanged,
\begin{align}\label{eq:constraintsA}
\int \dd \xvu_B |\Psi_R(\xvu_A, \xvu_B)|^2 &= |\Psi_0^A(\xvu_A)|^2 ,\\
\int \dd \xvu_A |\Psi_R(\xvu_A, \xvu_B)|^2 &= |\Psi_0^B(\xvu_B)|^2. \label{eq:constraintsB}
\end{align}
The wavefunction $\Psi_R$ in eq~\eqref{eq:EdispLevy} is searched over all wavefunctions antisymmetric only in the $\xvu_A$ and $\xvu_B$ variables separately, so that we are not considering the energy terms vanishing exponentially with $R$ due to the antisymmetrization between $A$ and $B$. That is, we work within the polarization approximation, as usual with dispersion.
Equation~\eqref{eq:EdispLevy} can also be written in the simpler form
\begin{equation}\label{eq:Levysimpler}
	 E_{\rm disp}^\method(R)=\min_{\Psi_R\to \diamatA,\diamatB}\langle\Psi_R|\hat{T}+\hat{V}_{ee}^{AB}|\Psi_R\rangle-T_0^A-T_0^B-U[\rho^A,\rho^B],
\end{equation}
with $T_0^{A/B}$ the ground-state kinetic energies of the two separated systems and 
\begin{equation}
U[\rho^A,\rho^B]=\int d\rv\int d\rv'\frac{\rho^A(\rv)\rho^B(\rv')}{|\rv-\rv'|},
\end{equation}
with $\rho^{A/B}(\rv)$ the monomer ground-state densities corresponding to $|\Psi_0^{A/B}|^2$.
The constraints of eqs~\eqref{eq:constraintsA}-\eqref{eq:constraintsB} ensure that $E_{\rm disp}^\method$ is a variational approximation for the interaction energy in the polarization approximation, since all the intramonomer potential energy terms (electrons-nuclei and electrons-electrons) cancel out exactly in $E_{\rm disp}^\method$ as they are fully determined by $\diamatA$ and $\diamatB$. Thus
\begin{equation}
	E_{\rm disp}^\method(R) + U[\rho^A,\rho^B] + V_{\mathrm{ext},A}[\rho_B] +V_{\mathrm{ext},B}[\rho_A] \ge E_{\rm pol}(R),
\end{equation}
where we denote with $V_{\mathrm{ext},A/B}[\rho_{B/A}]$ the interaction of the electrons in B/A with the external potential of the nuclei in A/B. The interaction energy within the polarization approximation is defined as,
\begin{equation}
E_\mathrm{pol}(R) = E_{AB, \mathrm{pol}}(R)- E_0^A - E_0^B,
\end{equation}
where $E_0^{A/B}$ are the ground-state energies of the two separated systems and $E_{AB, \mathrm{pol}}(R)$ is the ground-state energy of the combined systems within the polarization approximation. The physics behind $E_{\rm disp}^\method(R)$ is very simple: the correlation of the electrons in $A$ with those in $B$ lowers the expectation value of  $\hat{V}_{ee}^{AB}$ while increasing the kinetic energy. By eliminating the role of the intrafragment potential energy, dispersion is reduced here to a competition between these two effects, providing a microscopic mechanism using a ground-state formalism. However, we should immediately remark that this is not what happens in the exact case, where the intramonomer potential energy terms are not unchanged but {\it increase} and the kinetic energy {\it decreases}. This aspect will be fully analysed in sec~\ref{sec:thm}, where it will be shown how the intramonomer energy is reshuffled between potential and kinetic energy to still yield the exact result for any two one-elecron monomers, and still be rather accurate for closed-shell atoms and molecules as we reported in refs~\citenum{KooGor-JPCL-19,KooWecGor-arxiv-20}.  

We should also mention that fixing the diagonal of the full many-body density matrix of each monomer is not the minimal variational restriction needed to make the intrafragment potential energy terms cancel out in the interaction energy. Keeping only the pair densities $P_2^{A/B}(\rv,\rv')$ of the two monomers unchanged would be enough. The fixed-pair densities (FPD) dispersion energy, 
\begin{equation}
E_{\rm disp}^{\rm FPD}(R) =	\min_{\Psi_R\to P_2^A,P_2^B}\langle\Psi_R|\hat{T}+\hat{V}_{ee}^{AB}|\Psi_R\rangle-T_0^A-T_0^B-U[\rho^A,\rho^B],
\end{equation}
would have more variational freedom, implying $E_{\rm disp}^\method(R)\ge E_{\rm disp}^{\rm FPD}(R)$. However, as we will detail below, the construction keeping the full many-body diagonal fixed takes a rather simplified form, leading in a natural way to the dispersion energy as a functional of the ground-state pair densities (or density and xc holes) of the two fragments.

\subsection{Variational ansatz for the $\method$ wavefunction}
A variational ansatz for the minimising wavefunction $\Psi_R$ in eq~\eqref{eq:Levysimpler} can be explicitly constructed following our recent work,\cite{KooGor-JPCL-19} which we review here, rewriting it and analysing it in terms of DFT quantities. The ansatz reads as
\begin{equation}\label{eq:PsiR}
\Psi_R(\xvu_A, \xvu_B) = \Psi_0^A(\xvu_A) \Psi_0^B(\xvu_B) \sqrt{1+\sum_{i\in A, j \in B}J_R(\rv_i, \rv_j)},
\end{equation}
where, as long as we deal with the situation at large $R$, we have $|J_R|\ll 1$. In fact, to leading order we find variationally that $J_R\propto R^{-3}$. The Jastrow-like function $J_R$ is expanded in a finite set of ``dispersals'' $b_i(\rv)$, introducing a set of variational parameters $c_{ij, R}$,
\begin{equation}
J_R(\rv_1, \rv_2) = \sum_{ij} c_{ij, R} b_i^A(\rv_1) b_j^B(\rv_2).
\end{equation}
The many-body density constraints of eqs~\eqref{eq:constraintsA}-\eqref{eq:constraintsB},
can be easily enforced by imposing the conditions,
\begin{align}
\int \dd \rv \,b_i^A(\rv) \rho^A(\rv) &= 0 \,\, \forall \, i,\\
\int \dd \rv\, b_j^B(\rv) \rho^B(\rv) &= 0 \,\, \forall \, j,
\end{align}
which reduces to a Gram-Schmidt orthogonalization w.r.t.~the electron density, rewriting the dispersals $b_i(\rv)$ in the form,
\begin{equation}
b_i(\rv) = f_i(\rv) - \frac{1}{N} \int \dd \rv' \rho(\rv') f_i(\rv') =: f_i(\rv) - p_i,
\end{equation}
where $f_i(\rv)$ is the set of dispersals before the Gram-Schmidt orthogonalization. 

We should remark that the ansatz of eq~\eqref{eq:PsiR} does not exhaust the space of all possible $\method$ wavefunctions. As such, it yields an upper bound to the $\method$ dispersion energy defined in eq~\eqref{eq:Levysimpler}. The ansatz is exact for any pair $A$ and $B$ of one-electron systems,\cite{KooGor-JPCL-19} and rather accurate for closed-shell many-electron systems.\cite{KooWecGor-arxiv-20} In a loose way, we denote in the following the interaction energy obtained by a variational optimisation on the class of wavefunctions \eqref{eq:PsiR} as $\method$ dispersion energy, although, strictly speaking, it is only a variational upper bound to it.

\subsection{Intermonomer pair density and xc-hole projection}
The ansatz for $\Psi_R$ corresponds to the following intermonomer dispersion pair-density $P^{AB}_{\mathrm{disp}}(\rv_A,\rv_B)=P^{AB}(\rv_A,\rv_B)-\rho^A(\rv_A)\rho^B(\rv_B)$
\begin{equation}\label{eq:Pdisp}
	P^{AB}_{\mathrm{disp}}(\rv_A,\rv_B)=\rho^A(\rv_{A}) \rho^B(\rv_{B})\sum_{ij}c_{ij,R}\,\Delta b_{i,xc}^A(\rv_{A})\,\Delta b_{j,xc}^B(\rv_{B}),
\end{equation}
where the difference $\Delta b_{i,xc}^{A/B}(\rv)$ between each dispersals $b_i^{A/B}(\rv)$ and its exchange-correlation (xc) hole projection $b_{i,xc}^{A/B}(\rv)$ is
\begin{subequations}
	\begin{align}
	\Delta b_{i,xc}^{A/B}(\rv) & = b_i^{A/B}(\rv)-b_{i,xc}^{A/B}(\rv), \\
	b_{i,xc}^{A/B}(\rv) & = -\int h_{xc}^{A/B}(\rv,\rv')\,b_i^{A/B}(\rv')\, d\rv',
	\end{align}
\end{subequations}
with the usual definition of the xc-hole
\begin{equation}
h_\mathrm{xc}^{A/B}(\rv_1, \rv_2) = \frac{P_2^{A/B}(\rv_1, \rv_2)}{\rho^{A/B}(\rv_1)}-\rho^{A/B}(\rv_2),
\end{equation}
where $P_2^{A/B}(\rv_1, \rv_2)$ is the ground-state pair density of each isolated monomer, normalised to $N_{A/B}(N_{A/B}-1)$. 

The properties and the idea of the xc-hole projection were studied in ref~\citenum{GorAngSav-CJC-09}, where for any multiplicative monoelectron operator $\hat{O}=\sum_i o(\rv_i)$ an associated xc-hole projected operator $\hat{O}_{xc}=\sum_i o_{xc}(\rv_i)$ was defined, with
\begin{equation}
o_{xc}(\rv)=-\int h_{xc}(\rv,\rv')\,o(\rv')\, d\rv'.
\end{equation}
By virtue of the sum rules obeyed by the xc-hole (simply due to the normalisation of the pair density),
\begin{subequations}
	\begin{align}
		\int d\rv'\,h_{xc}(\rv,\rv') & =-1 \label{eq:sumrulesimple} \\
		\int d\rv\,\rho(\rv)\,h_{xc}(\rv,\rv') & = -\rho(\rv'), \label{eq:sumruledifficult}
	\end{align}
\end{subequations}
we have that $\hat{O}$ and $\hat{O}_{xc}$ have the same ground-state expectation value,
\begin{equation}\label{eq:xcprojExp}
\langle\Psi|\hat{O}|\Psi\rangle=\langle\Psi|\hat{O}_{xc}|\Psi\rangle=\int d\rv\,\rho(\rv) o(\rv)=\int d\rv\,\rho(\rv) o_{xc}(\rv).
\end{equation}
For example, if $o(\rv)=\rv$, then $\rv_{xc}(\rv)$ is the position of the xc-hole barycenter of charge. In this case, eq~\eqref{eq:xcprojExp} simply shows that the expectation value of the electronic dipole moment can also be computed as a weighted sum of the xc-hole barycenters, which can be regarded\cite{GorAngSav-CJC-09} as a generalisation of the Wannier-center decomposition of the polarization.\cite{MarVan-PRB-97} 

Clearly, it always holds that $\Delta o_{xc}(\rv)=o(\rv)-o_{xc}(\rv)$ has zero expectation value for any monoelectron local operator. In our case, the $b_i(\rv)$ dispersals (and thus their associated $b_{i,xc}$) have also zero expectation value separately, by construction.
It is then also evident that $P^{AB}_{\mathrm{disp}}(\rv_A,\rv_B)$ of eq~\eqref{eq:Pdisp} has zero marginals (as it should, since it is not allowed to alter the ground-state monomer densities),
\begin{equation}\label{eq:PdispZeroMarginal}
\int P^{AB}_{\mathrm{disp}}(\rv_A,\rv_B)\,d\rv_A=\int P^{AB}_{\mathrm{disp}}(\rv_A,\rv_B)\,d\rv_B=0.
\end{equation}

\subsection{The $\method$ dispersion energy}
The dispersion energy $E_\mathrm{disp}^\method(R)$ can then be computed from our ansatz waverfunction in terms of the variational parameters $c_{ij,R}$, whose optimisation takes a simple form, as we briefly review here.\cite{KooGor-JPCL-19,KooWecGor-arxiv-20} We first rewrite eq~\eqref{eq:Levysimpler} as
\begin{equation}\label{eq:optimisEdisp}
E^\method_{\rm disp}[\rho^A, \rho^B, h_\mathrm{xc}^A, h_\mathrm{xc}^B](R)  =\min_{c_{ij,R}}\, \tilde{E}_\mathrm{disp}^\method[\rho^A, \rho^B, h_\mathrm{xc}^A, h_\mathrm{xc}^B](\{ c_{ij,R} \}), 
\end{equation}
where the expression $\tilde{E}_\mathrm{disp}^\method[\rho^A, \rho^B, h_\mathrm{xc}^A, h_\mathrm{xc}^B](\{ c_{ij,R} \})$ is computed from the expectation value on $\Psi_R$ as follows.
The expectation value of $V_{ee}^{AB}$ is just an integral with the intermonomer pair density $P^{AB}_{\mathrm{disp}}(\rv_A,\rv_B)$ of eq~\eqref{eq:Pdisp} and it is thus {\it linear} in the variational parameters. The kinetic energy is slightly more involved, but can be expanded through second order in the $c_{ij,R}$, which are small by construction (see the supplemental material of ref~\citenum{KooGor-JPCL-19} for the full derivation), leading to a {\it quadratic}  dependence on the $c_{ij,R}$. We should stress that the linear term in the kinetic energy disappears only by virtue of the fixed-diagonal density matrices constraint.\cite{KooGor-JPCL-19} 
The energy  $\tilde{E}_\mathrm{disp}^\method(\{ c_{ij,R} \})$ through second order in the $c_{ij,R}$ (denoted below as simply $c_{ij}$ for convenience) thus takes the form
\begin{equation}
\tilde{E}_\mathrm{disp}^\method[\rho^A, \rho^B, h_\mathrm{xc}^A, h_\mathrm{xc}^B](\{ c_{ij} \}) = \sum_{ij} c_{ij} w_{ij} + \frac{1}{8} \sum_{ijkl}c_{ij} c_{kl}\left(\tau_{ik}^A S_{jl}^B + S_{ik}^A \tau_{jl}^B\right), \label{eq:energy}
\end{equation}
where,
\begin{subequations}
\label{eq:matrices}
\begin{align}
w_{ij} =& \int \dd \rv_{A} \dd \rv_{B}\,w_\mathrm{disp}(\rv_{A}, \rv_{B})\, \rho^A(\rv_{A})\rho^B(\rv_{B})\Delta b_{i,xc}^A(\rv_{A})\,\Delta b_{j,xc}^B(\rv_{B})
 \label{eq:wij}\\
\tau_{ij}^{A/B} =& \int \dd \rv \,\rho^{A/B}(\rv)\, \nabla b_i^{A/B}(\rv) \cdot \nabla b_j^{A/B}(\rv), \label{eq:taui} \\
S_{ij}^{A/B} =& \int \dd \rv\, \rho^{A/B}(\rv) b_i^{A/B}(\rv)\,\Delta b_{j,xc}^{A/B}(\rv). \label{eq:Sij}
\end{align}
\end{subequations}
In eq~\eqref{eq:wij} the intermonomer interaction $w_\mathrm{disp}(\rv_{A}, \rv_{B})$ is simply $1/|\rv_A-\rv_B|$ or, equivalently, $1/|\rv_A-\rv_B|-U[\rho^A,\rho^B]$. In fact, by virtue of eq~\eqref{eq:PdispZeroMarginal}, any term in the interaction which does not contain both variables $\rv_A$ and $\rv_B$ has zero expectation value, so that both electrostatics and induction do not appear in $E_\mathrm{disp}^\method$.\\
\\
 With respect to our previous work\cite{KooGor-JPCL-19} there are some differences in notation: we are now denoting with  $S$ the quantity that was refered to as $S+P$. Although not immediately evident, the matrix  $S$ of eq~\eqref{eq:Sij} is symmetric by virtue of the xc-hole definition. Furthermore, the expression here is written in terms of the exchange-correlation hole projected dispersals instead of the isolated monomer pair densities. Aside from bringing it in line with the notation commonly used in DFT, the expressions of eq~\eqref{eq:wij}-\eqref{eq:Sij}  have the advantage of being immediately invariant under a constant shift of the dispersals. More explicitly, if one substitutes $b_i(\rv) = f_i(\rv) - p_i$, then the equations \eqref{eq:matrices} are simply those with $b_i(\rv)$ substituted by $f_i(\rv)$. \\
 Since the energy in equation \eqref{eq:energy} is quadratic and we have no additional constraints, we can obtain $c_{ij}$ by solving a linear system. However, the exact solution of this linear system scales as $\mathcal{O}(n_A^3 n_B^3)$, where $n_{A/B}$ is the number of functions $b_i^{A/B}$ included in the calculation. This can be reduced to $\mathcal{O}(n_A^3+ n_B^3)$ for every pair of systems $A$ and $B$ by diagonalizing the matrix $S$ using e.g. a L\"owdin orthogonalization. In that case, the solution of eq~\eqref{eq:optimisEdisp}
results in the solution of a Sylvester equation as described in ref~\citenum{KooGor-JPCL-19},
\begin{equation}
\frac{1}{4} \tau^A c + \frac{1}{4} c \tau^B = - w.
\end{equation}
 A further simplification can be made by solving instead the generalized eigenvalue problem for the matrix $\tau_{ij}^A$ with $S_{ij}^A$ as a metric (and similar for system $B$).\cite{KooWecGor-arxiv-20} In this case, we obtain directly a solution to the minimization, which takes the simplified form (where from now on indices $ij$ indicates the transformed dispersals)
 \begin{equation}
 c_{ij} = -\frac{4 w_{ij}}{\tau_i^A + \tau_j^B},
 \end{equation}
and leads to the $\method$ dispersion energy in terms of $\rho^A, \rho^B, h_\mathrm{xc}^A, h_\mathrm{xc}^B$, and as a function of the monomer-monomer distance $R$ which enters in $w_{ij}$
 \begin{equation}
 \label{eq:finalenergy}
 E_\mathrm{disp}^\method[\rho^A, \rho^B, h_\mathrm{xc}^A, h_\mathrm{xc}^B](R) = -2 \sum_{ij}\frac{w_{ij}^2}{\tau_i^A + \tau_j^B}.
 \end{equation}
The computational scaling is then still $\mathcal{O}(n_A^3+n_B^3)$ due to the diagonalisation of $\tau_{ij}^{A/B}$, but we only need to perform the diagonalisation once for each system instead of solving it for every pair of systems.  

Equation~\eqref{eq:finalenergy} provides the dispersion energy as a function of the monomer-monomer distance $R$ and their orientation. If we want to compute dispersion coefficients, we need to expand the interaction between the two systems in terms of multipoles,
\begin{equation}
w_\mathrm{disp}(\rv_{1_A}, \rv_{1_B}) = \sum_{n=3}^\infty \frac{w_{\mathrm{int}}^{(n)}(\rv_{1_A}, \rv_{1_B})}{R^n},
\end{equation}
where $w_{\mathrm{int}}^{(n)}(\rv_{1_A}, \rv_{1_B})$ is a sum of separable terms in $\rv_{1_A}$ and $\rv_{1_B}$, which, due to the linearity of $w_{ij}$ in the $c_{ij,R}$, leads to an expansion of the coefficients $c_{ij,R}$ of the form,
\begin{equation}
c_{ij,R} = \sum_{n=3}^\infty \frac{c_{ij}^{(n)}}{R^n}.
\end{equation}
The energy is then given as a power series in orders of $\frac{1}{R}$ as,
\begin{equation}
E_\mathrm{disp}^\method[\rho^A, \rho^B, h_\mathrm{xc}^A, h_\mathrm{xc}^B](R) = - \sum_{m, n=3}^\infty\frac{2}{R^{m+n}} \sum_{ij} \frac{w_{ij}^{(n)}w_{ij}^{(m)}}{\tau_i^A+\tau_j^B},
\end{equation}
where $w_{ij}^{(n)}$ is defined similarly to $w_{ij}$ in equation \eqref{eq:wij}, but with $w_\mathrm{int}^{(n)}$ instead of $w_\mathrm{int}$, thus becoming products of single-monomer integrals. The corresponding expressions for the dispersion coefficients are
\begin{equation}
C_n^\method[\rho^A, \rho^B, h_\mathrm{xc}^A, h_\mathrm{xc}^B] = \sum_{p+q=n} 2 \sum_{ij} \frac{w_{ij}^{(p)}w_{ij}^{(q)}}{\tau_i^A+\tau_j^B}.
\end{equation}
Notice that these $C_n^\method$ are orientiation-dependent through the $w_{ij}^{(p/q)}$. 

\subsection{$\method$ accuracy for the dispersion coefficients}
Before considering to make approximations for the xc holes of the monomers, one should ask the question: how accurate can the $\method$ dispersion energy be if we use the exact densities and xc holes of the monomers? In other words: Is the variational freedom (with fixed diagonal density matrices) too small to give accurate results even if we treat the monomers exactly?
In ref~\citenum{KooGor-JPCL-19} the $C_n^\method$ coefficients have been computed for the H-H case, using simple multipoles for the dispersal functions $b_i(\rv)$. It has been found that the $\method$ results yields the {\em exact} $C_6$, $C_8$ and $C_{10}$, and the exact second-order results for all the even $C_n$ up to $C_{30}$, with a fast convergence with the number of dispersals $b_i(\rv).$  Results for the He-He and He-H cases have a very small error, $\sim 0.17\%$, when using the very accurate He xc-hole of ref~\citenum{FreHuxMor-PRA-84}.

In ref~\citenum{KooWecGor-arxiv-20}, $C_6^\method$ coefficients have been computed for 459 pairs of atoms, ions and molecules using Hartree-Fock, MP2 and CCSD densities and xc-holes for the monomers, using again simple multipoles for the dispersals $b_i(\rv)$, 
\begin{equation}\label{eq:bi}
	b_i(\mathbf{r}) = (x-x_0)^{s_i} (y-y_0)^{t_i} (z-z_0)^{u_i},
\end{equation} 
centred in $\mathbf{r}_0 =(x_0,\, y_0,\, z_0)$, fixed at the barycenter of nuclear mass.
For closed-shell atoms and molecules, the isotropic $C_6^\method$ coefficients using CCSD xc-holes and densities for the monomers have errors around 7\%, with little sensitivity to the basis set used for the monomer calculations. Anisotropies have a very similar accuracy. The results for open-shell systens are, instead, less accurate. The $\method$ should thus provide a good basis to build real-space dispersion models for closed-shell systems. For open-shell systems some more understanding/improvement is needed.\\
Before discussing how to build approximations and how to find an optimal choice for the disperals $b_i(\rv)$, we clarify some theoretical aspects of the $\method$ framework.

\section{Hellmann-Feynman, Hohenberg-Kohn and virial theorem}
\label{sec:thm}
In the case of two hydrogen atoms something remarkable occurs: due to the lack of induction the $\method$ wavefunction gives the exact energy up to order $\mathcal{O}(R^{-10})$. This opens 
three theoretical questions.
\subsection{Hellmann-Feynman theorem}
The first question is how we can obtain a dispersion force without density distortion, despite Feynman's statement that ``it is the attraction
of each nucleus for the distorted charge distribution of its own electrons that gives the attractive $1/R^7$ force'', \cite{Fey-PR-39} and despite this statement has been proven for molecules by Hunt\cite{Hun-JCP-90} using the electrostatic Hellmann-Feynman theorem. This question can be answered by looking  at Steiner's 1973 paper,\cite{Ste-JCP-73} where he noted that the Hellmann-Feynman result depends on whether one performs the derivative with respect to the nuclear position in the original coordinates or in the coordinates in which the electrons are centred on their respective nuclei.
In the first case, one obtains the result by Feynmann and the $C_6$ coefficient in the force depends on the wavefunctions perturbed to first-order in the dipole-dipole and dipole-quadrupole interactions, as well as perturbed to second-order in both the dipole-dipole and dipole-quadrupole interactions. In the second case one obtains the usual result, where the $C_6$ coefficient depends only on the wavefunction perturbed to first-order in the dipole-dipole interaction (per the Wigner $2n+1$ rule). The connection to the density stems from the fact that the expressions in the first case depend only on the density distortion at order $R^{-7}$, while in the second case they depend only on the distortion of the interfragment pair density at order $R^{-3}$, which is exact in the $\method$ approach. Physically speaking, in the first case one computes the {\it force acting on the nucleus} while in the second case one computes the {\it force acting on the whole atom}. Since we are in the approximation of infinite nuclear mass, these two forces are exactly the same. So there should be no contradiction.
\subsection{Hohenberg-Kohn theorem}
The second question is if the Hohenberg-Kohn theorem is violated by having an electron density (namely, the sum of the two non-interacting H densities) which is different than the exact one, but yields an energy that is exact up to and including order $\mathcal{O}(R^{-10})$. As was found by Hirschfelder and Eliason,\cite{HirEli-JCP-67} the exact density change is of order $\mathcal{O}(R^{-6})$ for two hydrogen atoms, while the $\method$ density is equal to its zero-order value at all orders in $R^{-1}$ by construction. 
From the point of view of perturbation theory, this is in agreement with Wigner $2n+1$ rule: the $\method$ wavefunction can be exact up to $\mathcal{O}(R^{-5})$ for atoms and can thus yield the exact energy at $\mathcal{O}(R^{-10})$. 
\subsection{Virial theorem}
The third question is what happens with the virial theorem when using the variational $\method$ wavefunction. In the $\method$ framework there is by construction an increase in the electronic kinetic energy coupled with a decrease in the (potential) interaction energy, while the other two components of the potential energy, the intra-fragment electrons-nuclei and electron-electron interactions remain unchanged. We compare this to the exact case: from the virial theorem \cite{Lev-QC-00} we get for two neutral atoms separated by a large distance $R$,
\begin{align}
\langle T \rangle_R &= - E - R \dfrac{\mathrm{d} E}{\mathrm{d} R},\\
\langle V \rangle_R &= 2 E + R  \dfrac{\mathrm{d} E}{\mathrm{d} R},
\end{align}
such that we get for large $R$,
\begin{align}
E &= E_\infty - \frac{C_6}{R^6}+\mathcal{O}(R^{-8})\\
\langle T \rangle_R &=\langle T \rangle_\infty - \frac{5 C_6}{R^6}+\mathcal{O}(R^{-8}),\\
\langle V \rangle_R &= \langle V \rangle_\infty + \frac{4 C_6}{R^6}+\mathcal{O}(R^{-8}),
\end{align}
since $C_6$ is positive, we see immediately that the $\method$ wavefunction violates the virial theorem, because the virial theorem implies a \textbf{de}crease in the kinetic energy of the electrons coupled with an \textbf{in}crease in the total potential energy. To compare the situation to the standard approach via Rayleigh-Schr\"odinger Perturbation Theory we performed calculations for the hydrogen dimer using the $\method$ wavefunction and Hylleraas Variational Perturbation Theory (VPT) including only the first-order correction to the wavefunction. The contributions to the different components of the energy at order $R^{-6}$ were computed and are listed in table \ref{tab:virthm}. As is clear from the table, both Hylleraas VPT and the $\method$ wvaefunction violate the virial theorem \textit{when carried out to second-order}, but produce the correct $C_6$ to numerical precision. Furthermore, in both cases and in the exact case the ratio between the zeroth-order Hamiltonian and the perturbation $V_\mathrm{int}$ is $+1:-2$. To obtain the correct ratio in VPT from the virial theorem also inside each monomer, higher order contributions to the wavefunction need to be included, while in $\method$ this will not happen at any order by construction.

Dispersion between two systems in their ground state is a competition between a distortion of the fragments's ground-state (which raise the energy with respect to $E_0^A+E_0^B$) and the interfragment interaction. As proven by Lieb and Thirring,\cite{LieThi-PRA-86} the raise in energy due to the disortion of the fragment's ground-state can be always made quadratic with respect to a set of variational parameters, with the interfragment interaction being linear. In the $\method$ approach, we force the quadratic raise in energy of the isolated fragments to be of kinetic energy origin only. In the case of two H atoms this still leads to the exact overall raise in energy of the monomers, as shown in table  \ref{tab:virthm}.

\begin{table}
\centering
\begin{tabular}{|c|c|c|c|}
\hline 
Component ( $C_6$) & Exact  & $\method$ & Hylleraas VPT\\
\hline
$V_{\rm int}$ & $-2$ & $-2$ & $-2$\\
$V_{\rm ext}$ & $+6$ & $0$ & $\approx 1.304$\\
$V= V_{\rm int}+V_{\rm ext}$ & $+4$  & $-2$ & $\approx -0.696$\\
$T$  & $-5$ & $+1$ & $\approx-0.304$\\
$H_0 = T+V_{\rm  ext}$ & $+1$ & $+1$ & $+1$\\ 
Total & $-1$ & $-1$ & $-1$\\
\hline
\end{tabular}
\caption{Components of the $C_6$ energy for the hydrogen dimer. The intrafragment terms $V_\mathrm{ext}$ and $T$ are the sum of the contribution from both the identical fragments, while $V_{\rm int}$ is the contribution from the interfragment interaction.}
\label{tab:virthm}
\end{table}

\section{Pure density functional and functional derivative}

\begin{table}
\centering
\begin{tabular}{|c|c|c|c|c|}
\hline 
& Accurate\cite{YanBabDal-PRA-96} & Physical & KS & SCE\cite{SeiGorSav-PRA-07}\\
\hline
$C_6$ (a.u.) & 1.460978 & 1.458440 & 1.70615 & 0.478433\\
\% of accurate & & 99.8\% & 116.8\% & 32.7\%\\
\hline
\end{tabular}
\caption{$C_6$ energies obtained for the Helium dimer using the electron density from the wavefunction of Freund, et al.\cite{FreHuxMor-PRA-84} and different exchange(-correlation) holes corresponding to the same density.}
\label{tab:xchole}
\end{table}
For given dispersals $b_i^{A/B}(\rv)$ (whose optimal choice will be discussed in the next sections), if we want to
transform the $\method$ model into a ``pure'' density functional method, we need to include a functional for the exchange-correlation hole of the monomers. One immediate approximation that comes to mind is the Kohn-Sham (KS) exchange hole of the monomers instead of the full exchange-correlation hole, a choice that reminds the XDM idea of Becke and Johnson.\cite{BecJoh-JCP-07} In table \ref{tab:xchole} we have used the KS, exact and Strictly-Correlated Electrons (SCE) exchange-correlation holes to compute the $C_6^\method$ coefficient for the He-He case. 
We see that the KS exchange hole alone overestimates $C_6$. This same trend was observed in ref~\citenum{KooWecGor-arxiv-20}: for both atoms and molecules, $C_6^\method$ with monomers' Hartree-Fock holes and densities are typically overestimated (errors around 50\%), and even more so when using exchange holes with KS orbitals from semilocal functionals. This suggests that some alternative approximation for the exchange-correlation hole, possibly directly targeting the  xc-hole projected dispersals, must be used. The SCE xc-hole heavily underestimates the $C_6$ coefficients (leading to a too high dispersion energy). An interpolation between KS and SCE could be also considered in future work, also using the MP2 information as in ref.~\citenum{VucIroSavTeaGor-JCTC-16}.

We should also remark that we need an approximate xc hole that satisfies not only the sum rule of eq~\eqref{eq:sumrulesimple}, but also the one of eq~\eqref{eq:sumruledifficult}. While all available approximations satisfy \eqref{eq:sumrulesimple}, the second sum rule, eq~\eqref{eq:sumruledifficult}, is violated by all available semilocal xc-hole functionals, including the Becke-Roussel one,\cite{BecRou-PRA-89} as discussed in ref~\citenum{GorAngSav-CJC-09}. We should probably rather build simplified weighted-density-approximation (WDA) xc holes targeting accurate xc-hole projected dispersals.

If we have a reasonable xc-hole density functional and we want to include the effect of dispersion in a self-consistent manner, we would need the functional derivative of $E_\mathrm{disp}^\method[\rho^A, \rho^B]$, which can be obtained from eq~\eqref{eq:finalenergy}. This is straightforward but lengthy, since the $b_i(\rv)$ which diagonalize both $\tau$ and $S$ change when the density changes. Therefore, we only report the result and leave the full derivation in the supplementary material,
  \begin{align}
 \frac{\delta E_\mathrm{disp}^\method[\rho^A, \rho^B]}{\delta \rho^A(\mathbf{r})} =& 2 \sum_{pqr} \left(\dot{\tau}^A_{rp}(\rv)+\tau_q^B \dot{S}^A_{rp}(\rv) \right)\nonumber \frac{  w_{pq} w_{rq} }{(\tau_p^A + \tau_q^B) (\tau_r^A + \tau_q^B) }-4 \sum_{pq} \frac{  \dot{w}_{pq}^A(\rv) w_{pq}}{\tau_p^A + \tau_q^B}, 
 \end{align}
 where $\dot{\tau}^A_{rp}(\rv)$, $\dot{S}^A_{rp}(\rv)$ and $\dot{w}_{pq}^A(\rv)$ are the functional derivatives of the corresponding matrices in the ``atomic'' dispersal basis and transformed to the basis in which $\tau$ and $S$ are diagonal. Their explicit expressions can be found in the supplementary material.
 
\section{Choice of the dispersals}
The choice for the dispersals $b_i^{A/B}(\rv)$ of eq~\eqref{eq:bi} was suggested by the physics of dispersion interactions and by the immediate availability of matrix elements, but it is by no means optimal. In particular, it was observed in refs~\citenum{KooGor-JPCL-19} and \citenum{KooWecGor-arxiv-20} that, with this choice, convergence with the number of $b_i^{A/B}(\rv)$ functions varies a lot among systems. In some cases it can be quite fast, but in a few difficult cases (e.g., CS$_2$-CS$_2$), satisfactory convergence could not be really reached.\cite{KooWecGor-arxiv-20} In this section we discuss different strategies to optimise the choice of the dispersals. Before doing so, we discuss which functions are admissible. 
\subsection{Dispersal space}
Dispersals form a vector space with a weighted inner product,
 \begin{equation}
 \langle b_i, b_j \rangle_\rho = \int \dd \mathbf{r} \rho(\mathbf{r}) b_i(\rv) b_j(\rv)= \langle b_j, b_i \rangle_\rho),
 \end{equation}
 it is clear from equations \eqref{eq:matrices} that we must have,
 \begin{subequations}
 \begin{align}
 \tau_{ij} &= \langle \nabla b_i, \nabla b_j \rangle_\rho < \infty,\\
 S_{ij} &= \langle b_i, (1+h_\mathrm{xc}) b_j \rangle_\rho =  \langle (1+h_\mathrm{xc}) b_i,  b_j \rangle_\rho  < \infty,\\
 S & \succeq 0 
 \end{align}
 \end{subequations}
The overlap $S$ is somewhat unusual due to the presence of the exchange-correlation hole $h_\mathrm{xc}$. Indeed, while we are guaranteed that $\langle b_i | b_i \rangle_\rho > 0$, in general $\langle b_i | h_\mathrm{xc} b_i \rangle_\rho \leq 0$.  However, the inequality of Garrod and Percus\cite{Per-JCP-05,AyeDav-IJQC-06,GarPer-JMP-64} rewritten for the exchange-correlation hole,
 \begin{align}
  \int \dd \rv_1 \dd \rv_2 f(\rv_1) f(\rv_2) \rho(\rv_1) h_\mathrm{xc}(\rv_1, \rv_2) &\geq - \int \dd \rv f(\rv)^2 \rho(\rv)\,\, \forall \, f(\rv),
 \end{align}
implies directly $S_{ii} \geq 0$ by replacing $f(\rv)$ by $b_i(\rv)$. Then, via Cauchy-Schwarz we find,
 \begin{align}
 \tau_{ij}^2 &\leq \tau_{ii} \tau_{jj}\\
 S_{ij}^2 & \leq S_{ii} S_{jj},
 \end{align}
which means boundedness of the diagonal elements $S_{ii}$ and $\tau_{ii}$ implies boundedness of the off-diagonal elements.  Therefore we can define our space of dispersals as,
 \begin{equation}
 \mathcal{B}(\rho, h_\mathrm{xc}) = \{ b_i| 0 < S_{ii} < \infty, \tau_{ii} < \infty \}.
 \end{equation}
Note that the restriction $S_{ii} > 0$ rules out only functions that are constant everywhere. The dispersals, very much unlike orbitals, do not decay to zero as $\rv \rightarrow \infty$ for a finite system. All that seems to be necessary for valid dispersals is that $\lim_{\rv \rightarrow \infty} \rho(\rv) b_i(\rv)^2$ and $\lim_{\rv \rightarrow \infty} \rho(\rv) |\nabla b_i(\rv)|^2$ go to zero fast enough for the integrals to be finite. Since Coulombic densities decay exponentially and, therefore, have all finite moments, all non-constant polynomials are valid dispersals. It is at this moment, however, still unclear what a ``complete set'' of dispersals consist of.\\
\\
Finally, a comment on the finiteness of the remaining matrix elements $w_{ij}^{(n)}$ is in order. Since $w_{ij}^{(n)}$ is a finite sum of terms proportional to $\langle x^a y^b z^c, (1+h_\mathrm{xc})b_i^A \rangle_{\rho_A} \langle x^d y^e z^f, (1+h_\mathrm{xc}) b_j^B \rangle_{\rho_B} $, we also immediately obtain the result that $|w_{ij}^{(n)}| < \infty$, if $b_i \in  \mathcal{B}(\rho^A, h_\mathrm{xc}^A)$ and  $b_j \in  \mathcal{B}(\rho^B, h_\mathrm{xc}^B)$, and $\rho^A$ an $\rho^B$ decay exponentially or faster. This is not a restriction, since we are dealing with Coulombic systems, where the density always decays exponentially.

\subsection{Strategies to choose the best dispersals}

As mentioned, our previous calculations\cite{KooGor-JPCL-19,KooWecGor-arxiv-20} used the simple dispersals of eq~\eqref{eq:bi}, but to get a faster convergence we would like to use a more specialised set of ``atomic'' dispersals, which can then be used as a basis set for molecular calculations.  This is allowed, since if we perform an orthogonal transformation of the dispersals, $b_a^{A/B}(\rv) = \sum_i O_{ia}^{A/B} \tilde{b}_i^{A/B}(\rv)$, then the energy expression remains the same, however with transformed coefficients $c_{ab} = \sum_{ij} O_{ia}^{A} O_{jb}^{B} \tilde{c}_{ij}$. In the following we will discuss several approaches: diagonalisation of $\tau$, optimisation of a limited set of dispersals, and ``natural'' atomic dispersals.

\subsubsection{Diagonalising $\tau$}
Since we are keen on using dispersals that diagonalize $\tau$ with $S$ as a metric, one option might be to attempt to find the corresponding orbitals in real space, where $\hat{\tau}$ has the form
\begin{equation}
\hat{\tau} = - \nabla^2 - \frac{\nabla \rho(\mathbf{r})}{\rho(\mathbf{r})}  \cdot \nabla, \label{eq:tauspatial}
\end{equation}
while $\hat{S}$ is a non-local operator:
\begin{equation}
\hat{S}\, b(\rv) = b(\rv) + \int \dd \rv' h_\mathrm{xc}(\rv, \rv') b(\rv) \label{eq:Sspatial}.
\end{equation}
In general we cannot find the analytic solution to this problem, due to the presence of $h_\mathrm{xc}$. However, it is useful to examine the hydrogen atom case to get insight into this strategy (and rule it out, as we will see). In this case we can solve for $b_i$ exactly and obtain that $b_i(\rv)$ are related to the hydrogenic orbitals $\phi_{n_i,l_i,m_i}$,
\begin{align}
b_i(\rv) = \frac{\phi_{n_i,l_i,m_i}(\rv)}{\phi_0(\rv)},
\end{align}
with eigenvalues $\tau_i = \frac{n(n+2)}{(n+1)^2}$. This demonstrates directly that using the eigenstates of $\tau$ as a basis is not a viable route, since in this case it becomes equivalent to performing Rayleigh-Schr\"odinger perturbation theory with the hydrogenic orbitals, for which the convergence is notoriously slow and without the continuum one only obtains $\approx 60.2\%$ of $C_6$. In fact, in general for ground state one-electron systems we find that the dispersals diagonalising $\tau$ are related to the excited orbitals and we obtain a correspondence with second-order Rayleigh-Schr\"odinger perturbation theory, thus showing that the theory gives exactly the second-order dispersion energy for two one-electron systems.
\subsubsection{Optimising the dispersals}
An alternative approach to find optimal dispersals is to limit the dispersals to a small number $N_\mathrm{disp}$ and then optimize the dispersion energy w.r.t. the dispersals expressed in a larger basis. To this end, one minimizes the energy of eq~\eqref{eq:energy}, while enforcing the normalisation of the dispersals with a Lagrange multiplier $\lambda^{A/B}$ . The complete Lagrangian to optimize is then,
\begin{align}
L(\{O_{ia}^A, O_{ia}^B, c_{ab}, \lambda_{ac}^A, \lambda_{bd}^B\})  =& \sum_{ab}^{N_\mathrm{disp}}c_{ab} \sum_{ij} O_{ia}^A O_{jb}^B w_{ij} +\frac{1}{8} \sum_{acb}^{N_\mathrm{disp}} c_{ab} c_{cb} \sum_{ik} O_{ia}^A O_{kc}^A \tau_{ik}^A \nonumber\\
&+\frac{1}{8} \sum_{abd}^{N_\mathrm{disp}} c_{ab} c_{ad} \sum_{jl} O_{jb}^B O_{ld}^B \tau_{jl}^B - \sum_{ac}^{N_\mathrm{disp}} \lambda_{ac}^A (\sum_{ik} O_{ia}^A O_{kc}^A S_{ik}^A - \delta_{ac}) \nonumber \\ &- \sum_{bd}^{N_\mathrm{disp}} \lambda_{bd}^B (\sum_{jl} O_{jb}^B O_{ld}^B S_{jl}^B - \delta_{bd}).\label{eq:lagrangian}
\end{align}
Note that the first term is linear in the expansion coefficients $O^{A/B}$ of the dispersals of both systems, which makes the optimization w.r.t. $O$ different than in a Self-Consistent Field (SCF) procedure for Hartree-Fock or Kohn-Sham. Furthermore, when using more than one dispersal the coefficients $c$ need to be optimized like in a Multi-Configuration SCF (MCSCF) procedure.\\
\\
The optimisation can be carried out and the final dispersals can be fitted, but the procedure by itself yields little insight into properties such as asymptotic behaviour. To gain better understanding we analyze again the simple case of the interaction between two hydrogen atoms. In the case of two equal systems and only a single dispersal the Lagrangian of equation \eqref{eq:lagrangian} becomes,
\begin{align}
L(\{O_i, c\}) &= c \sum_{ij} O_i O_j w_{ij} + \frac{c^2}{4} \sum_{ij} O_i O_j  \tau_{ij} - \lambda (\sum_{ij} O_i O_j S_{ij} - 1) .
\end{align}
Optimising the Lagrangian w.r.t. $c$ yields,
\begin{align}
c &= -2 \frac{\sum_{ij} O_i O_j w_{ij}}{ \sum_{ij} O_i O_j \tau_{ij} } =: - 2 \frac{\langle w \rangle}{\langle \tau \rangle},\\
E_\mathrm{disp}[\rho^A, \rho^B] &= - \frac{\langle w \rangle^2 }{\langle \tau \rangle},
\end{align}
which transforms the Lagrangian to,
\begin{align}
L(\{O_i\}) &= -\frac{(\sum_{ij} O_i O_j w_{ij})^2}{ \sum_{ij} O_i O_j \tau_{ij} }  - \lambda (\sum_{ij} O_i O_j S_{ij} - 1).
\end{align}
Now we optimise w.r.t. $O_{i}$,
\begin{align}
\frac{\partial L(\{O_i\})}{\partial O_i} =& -2 \frac{\sum_{ij} O_i O_j w_{ij}}{ \sum_{ij} O_i O_j \tau_{ij} } \sum_j w_{ij}  O_j + \frac{(\sum_{ij} O_i O_j w_{ij})^2}{(\sum_{ij} O_i O_j \tau_{ij})^2} \sum_j \tau_{ij} O_j  \nonumber \\ &- \lambda \sum_{j} S_{ij}  O_j,  \nonumber \\
=& -2 \frac{\langle w \rangle}{ \langle \tau \rangle } \sum_j w_{ij}  O_j + \frac{\langle w \rangle^2}{\langle \tau \rangle^2} \sum_j \tau_{ij} O_j  - \lambda \sum_{j} S_{ij}  O_j .
\end{align}
We now solve the resulting equation for a given $\langle w \rangle$ and $\langle \tau \rangle$ from the previous iteration. That is, we solve the generalised eigenvalue problem,
\begin{equation}
- 2 \frac{\langle w \rangle^{(n-1)}}{\langle \tau \rangle^{(n-1)}} w O^{(n)} + \frac{\langle w \rangle^{(n-1)2}}{\langle \tau \rangle^{(n-1)2}} \tau O^{(n)} =S O ^{(n)} \lambda^{(n)} \label{eq:eigenvalue},
\end{equation}
for the eigenvalues $\lambda^{(n)}$ and eigenvectors $O^{(n)}$. Then we select the lowest (most negative) eigenvalue of $\lambda^{(n)}$ since at convergence we obtain (by left-multiplying equation \eqref{eq:eigenvalue} by $O_i$ and summing),
\begin{equation}
\lambda^{(n)} =  - \frac{\langle w \rangle^2}{\langle \tau \rangle} = E_\mathrm{disp}[\rho^A, \rho^B] ,
\end{equation}
which we want to minimise. This completes the specification of the iterative problem. 

\begin{figure}[h]
\centering
  \includegraphics[height=5.5cm]{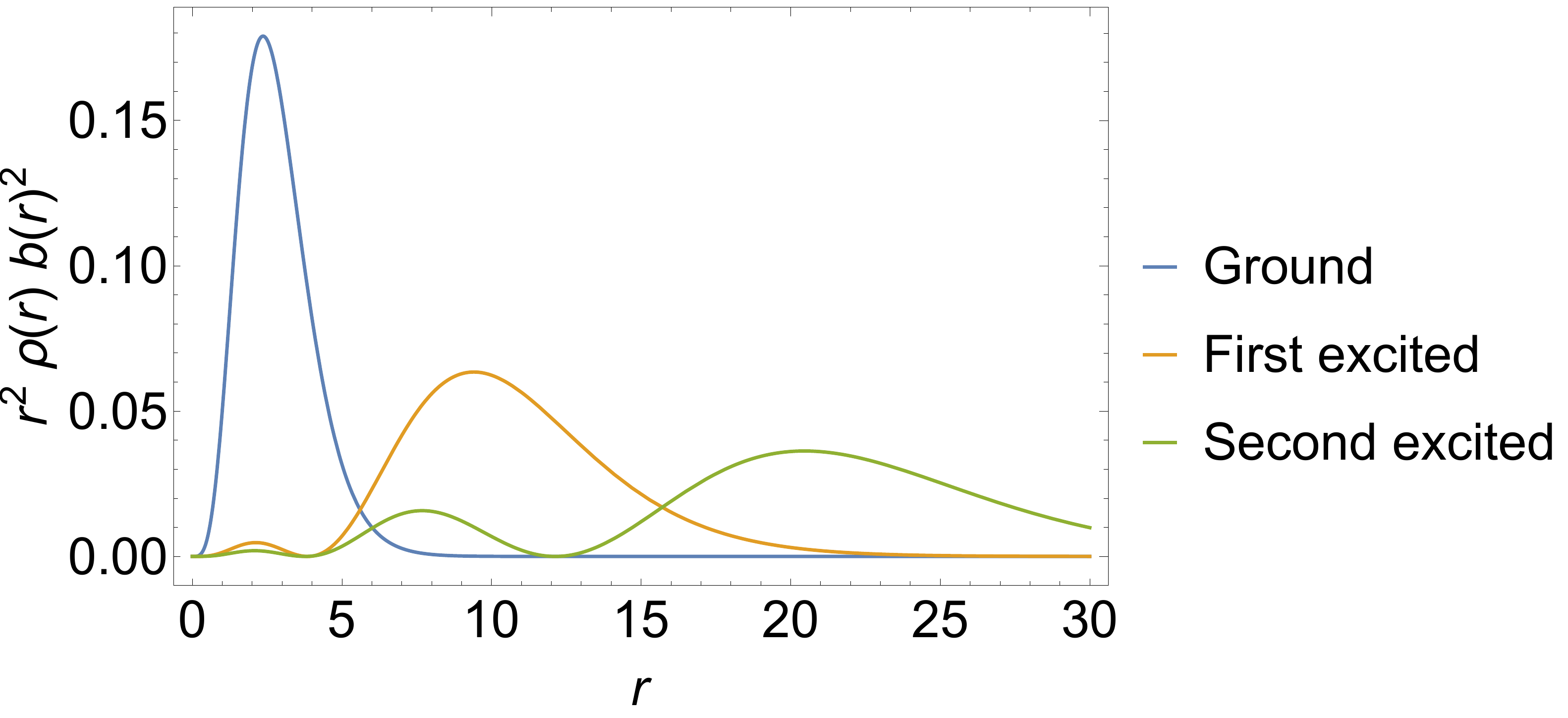}
  \caption{Lowest three eigenfunctions of equation \eqref{eq:eigenvalue} upon convergence of the iterative procedure. To be able to visualise all three solutions, we plot the square of the dispersal multiplied by the volume element and the density. }
  \label{fgr:iterativedispersaldens}
\end{figure}

We have performed the iterative procedure for the hydrogen 1s state with a basis consisting of $r^n$, for $n=1$ to $n=50$, in this case we recover, with a single radial dispersal, 99.97\% of the $C_6$ coefficient. In figure \ref{fgr:iterativedispersaldens} we plot the iterative solution, as well as the first and second ``excited states'' of equation \eqref{eq:eigenvalue}. The first and second excited states are much more diffuse,  in fact corresponding to \textit{positive} eigenvalues $\lambda$ and do not seem to be useful in a dispersal basis.

\begin{figure}[h]
\centering
  \includegraphics[height=5.5cm]{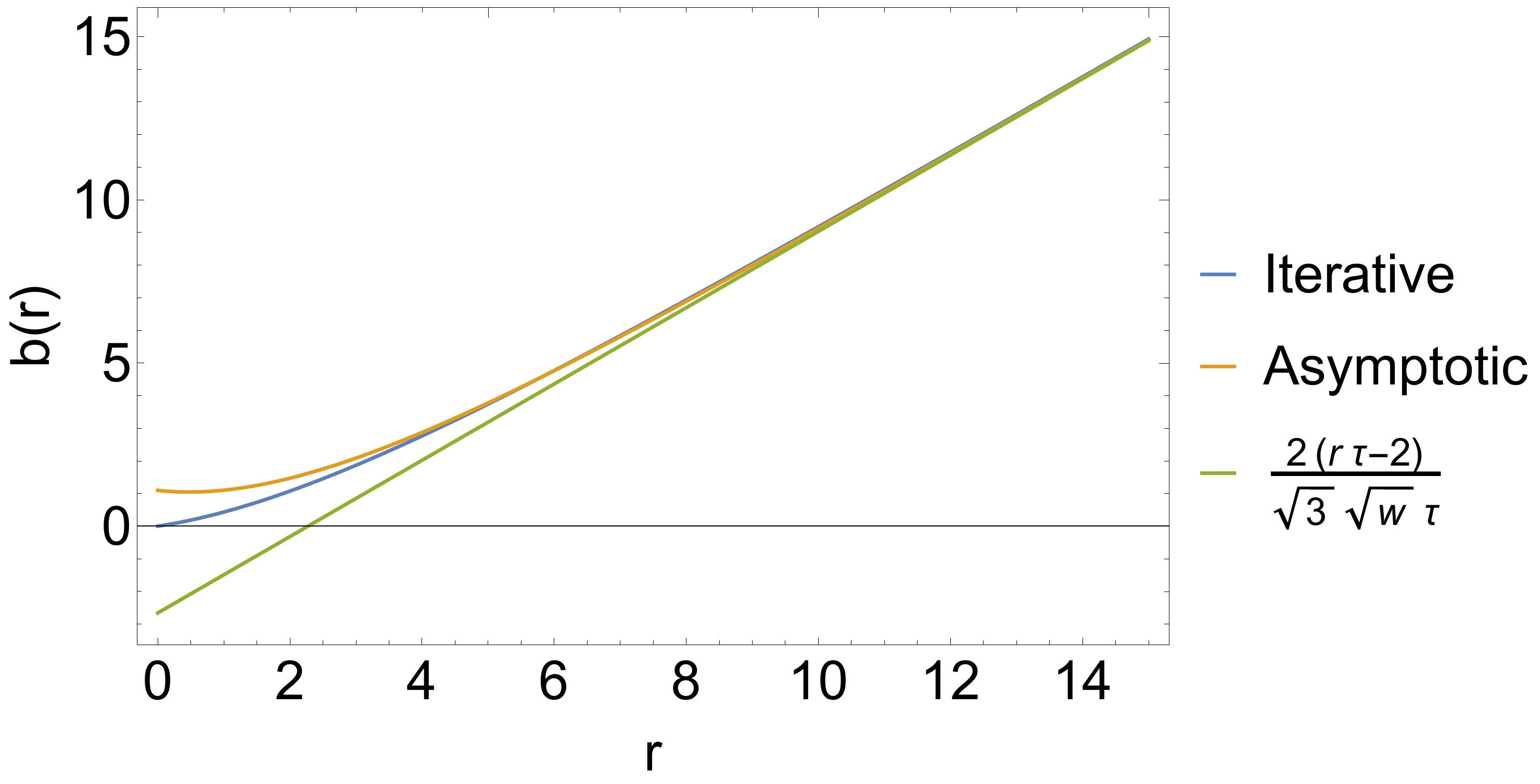}
  \caption{Iterative solution of equation \eqref{eq:eigenvalue} upon convergence of the iterative procedure compared to the analytical asymptotic behavior of equation \eqref{eq:asymp}.  For the asymptotic expression, we obtain from normalisation $c_1\approx3.756$.}
  \label{fgr:iterativedispersalasymp}
\end{figure}

 To obtain the asymptotic behaviour of $b(\rv)$ we will switch again to the spatial representation, which we have already obtained for $\hat{S}$ and $\hat{\tau}$ in equations \eqref{eq:Sspatial} and \eqref{eq:tauspatial}, respectively. From the functional derivative of $\hat{w}$ we obtain again a non-local operator,
\begin{equation}
\hat{w} b(\mathbf{r}) = \int \mathrm{d} \mathbf{r}' \rho(\mathbf{r}') b(\mathbf{r}') w(\mathbf{r}, \mathbf{r}').
\end{equation}
Now we move on to our specific case of the hydrogen atom, where $\rho(\mathbf{r}) = \frac{e^{-2r}}{\pi}$.
We have used three $b$-functions, because of the spherical symmetry, but all with the same radial part $b(r)$. We write,
\begin{equation}\label{eq:radialbDEF}
b_{l, m}(\mathbf{r}) = b(r)\,Y_{l}^m(\theta, \phi).
\end{equation}
Our $J$ function in this context is,
\begin{equation}
J(\mathbf{r}_1, \mathbf{r}_2) = \sum_{m=-1}^1 c_m b(r_1) b(r_2) Y_1^m(\theta_1, \phi_1) Y_1^{-m}(\theta_2, \phi_2).
\end{equation}
We specialise now to the dipole-dipole interaction. We have obtained that the coefficient in the $m=0$ direction is twice that of that in the $m=-1$ and $m=1$ directions.  We thus work with a single direction ($m=-1$ or $m=1$) and in the end multiply $\lambda$ by 6 to get the $C_6$ (see the supplementary material of our previous work\cite{KooGor-JPCL-19}). The interaction $w_\mathrm{int}(\mathbf{r}_1, \mathbf{r}_2)$ is given in spherical coordinates as,
\begin{align}
w_\mathrm{int}(\mathbf{r}_1, \mathbf{r}_2) =& -\frac{r_1 r_2 }{3} (2 Y_{1}^0(\theta_1, \phi_1) Y_{1}^0(\theta_2, \phi_2)+ Y_{1}^1(\theta_1, \phi_1) Y_{1}^{-1}(\theta_2, \phi_2) +Y_{1}^{-1}(\theta_1, \phi_1) Y_{1}^{1}(\theta_2, \phi_2)).
\end{align}
Letting $\hat{w}$ act on  $b_{-1}(\mathbf{r}_2) = Y_{1}^{-1}(\theta_2, \phi_2)b(r_2)$ we obtain,
\begin{equation}
\hat{w}\, b_{-1}(\mathbf{r}_1) = - r_1 Y_{1}^{-1}(\theta_1, \phi_1) \frac{1}{3}\int \mathrm{d} r_2\, r_2^3 \,b(r_2)\, \rho(r_2).
\end{equation}
Multiplying with $\rho(r_1) r_1^2 b(r_1) Y_{1}^{1}(\theta, \phi)$ and integrating we obtain $\langle w \rangle$,
\begin{equation}
\langle w \rangle =  - \frac{1}{3} \left(\int \mathrm{d} r\, r^3 \, b(r)\, \rho(r)\right)^2 .
\end{equation}
So we obtain as an expression for the action of $\hat{w}$ on $b_{-1}(\rv)$,
\begin{equation}
\hat{w}\, b_{-1}(\mathbf{r}_1) =  - r_1 Y_{1}^{-1}(\theta_1, \phi_1) \sqrt{\frac{-\langle w \rangle}{3}}.
\end{equation}
The spherical expression for $\tau$ acting on $b_{-1}(\rv)$ is, dividing by $ Y_1^{-1}$,
\begin{equation}
\frac{\hat{\tau}\, b_{-1}(\mathbf{r}_1)}{ Y_1^{-1}(\theta, \phi) } =- \frac{\partial^2 b(r)}{\partial r^2} - \frac{2}{r}\frac{\partial b(r)}{\partial r} - \frac{1}{\rho(r)} \frac{\partial \rho(r)}{\partial r} \frac{\partial b(r)}{\partial r} + \frac{2}{r^2} b(r),
\end{equation}
while the spatial expression for $\hat{S}$ is simply the identity. Putting everything together in the non-linear equation,
\begin{equation}
- 2 \frac{\langle w \rangle^{(n-1)}}{\langle \tau \rangle^{(n-1)}} \hat{w}\, b^{(n)}_{-1}(\rv) + \frac{\langle w \rangle^{(n-1)2}}{\langle \tau \rangle^{(n-1)2}} \hat{\tau}\, b^{(n)}_{-1}(\rv) = \lambda^{(n)}\, b^{(n)}_{-1}(\rv) \label{eq:eigenvalue2},
\end{equation}
and dividing out the spherical harmonic, we get the spherical differential equation,
\begin{align}
&\frac{-2 \langle w \rangle \sqrt{{\frac{|\langle w \rangle|}{l(l+1)}}}}{\langle \tau \rangle} r + \frac{\langle w \rangle^2}{\langle \tau \rangle ^2} \Big(- \frac{\partial^2 b(r)}{\partial r^2} - \frac{2}{r}\frac{\partial b(r)}{\partial r}  - \frac{1}{\rho(r)} \frac{\partial \rho(r)}{\partial r} \frac{\partial b(r)}{\partial r} \nonumber \\ & + \frac{l(l+1)}{r^2} b(r) \Big) = \lambda b(r) =  - \frac{\langle w \rangle^2}{\langle \tau \rangle} b(r).
\end{align}
In the case of the hydrogen atom, $\frac{1}{\rho(r)} \frac{\partial \rho(r)}{\partial r} = -2$,
\begin{align}
&\frac{-2 |\langle w \rangle|^{3/2}}{ \sqrt{3}\langle \tau \rangle} r + \frac{\langle w \rangle^2}{\langle \tau \rangle ^2}\Big(- \frac{\partial^2 b(r)}{\partial r^2} - \frac{2}{r}\frac{\partial b(r)}{\partial r} + 2 \frac{\partial b(r)}{\partial r} + \frac{2}{r^2} b(r) \Big), = \lambda b(r) =  - \frac{\langle w \rangle^2}{\langle \tau \rangle} b(r).
\end{align}
The solution of this inhomogeneous partial differential equation (PDE) is not trivial. However, to analyse the asymptotic behaviour at large $r$, we look at the leading $r \rightarrow \infty$ terms and solve the simpler problem,
\begin{align}
\frac{-2 |\langle w \rangle|^{3/2}}{ \sqrt{3}\langle \tau \rangle} r + \frac{\langle w \rangle^2}{\langle \tau \rangle ^2}\left(- \frac{\partial^2 b(r)}{\partial r^2} + 2 \frac{\partial b(r)}{\partial r} \right) = \lambda b(r) =  - \frac{\langle w \rangle^2}{\langle \tau \rangle} b(r),
\end{align}
finding that the asymptotic solution for $r\rightarrow \infty$ is 
\begin{equation}
b(r) \sim \frac{2 r}{\sqrt{3 \langle w \rangle}} - \frac{4}{\sqrt{3 \langle w \rangle} \langle \tau \rangle} + c_1 e^{r(1- \sqrt{1+\langle \tau \rangle})} + c_2 e^{r(1+ \sqrt{1+\langle \tau \rangle})}  \label{eq:asymp}.
\end{equation}
To ensure the dispersal be normalisable, we need that $c_2 = 0$. $c_1$ is then fixed by normalisation. We verify the asymptotic behaviour in figure \ref{fgr:iterativedispersalasymp}. 

\subsubsection{Natural dispersals}
\begin{figure}[h]
\centering
  \includegraphics[height=5.5cm]{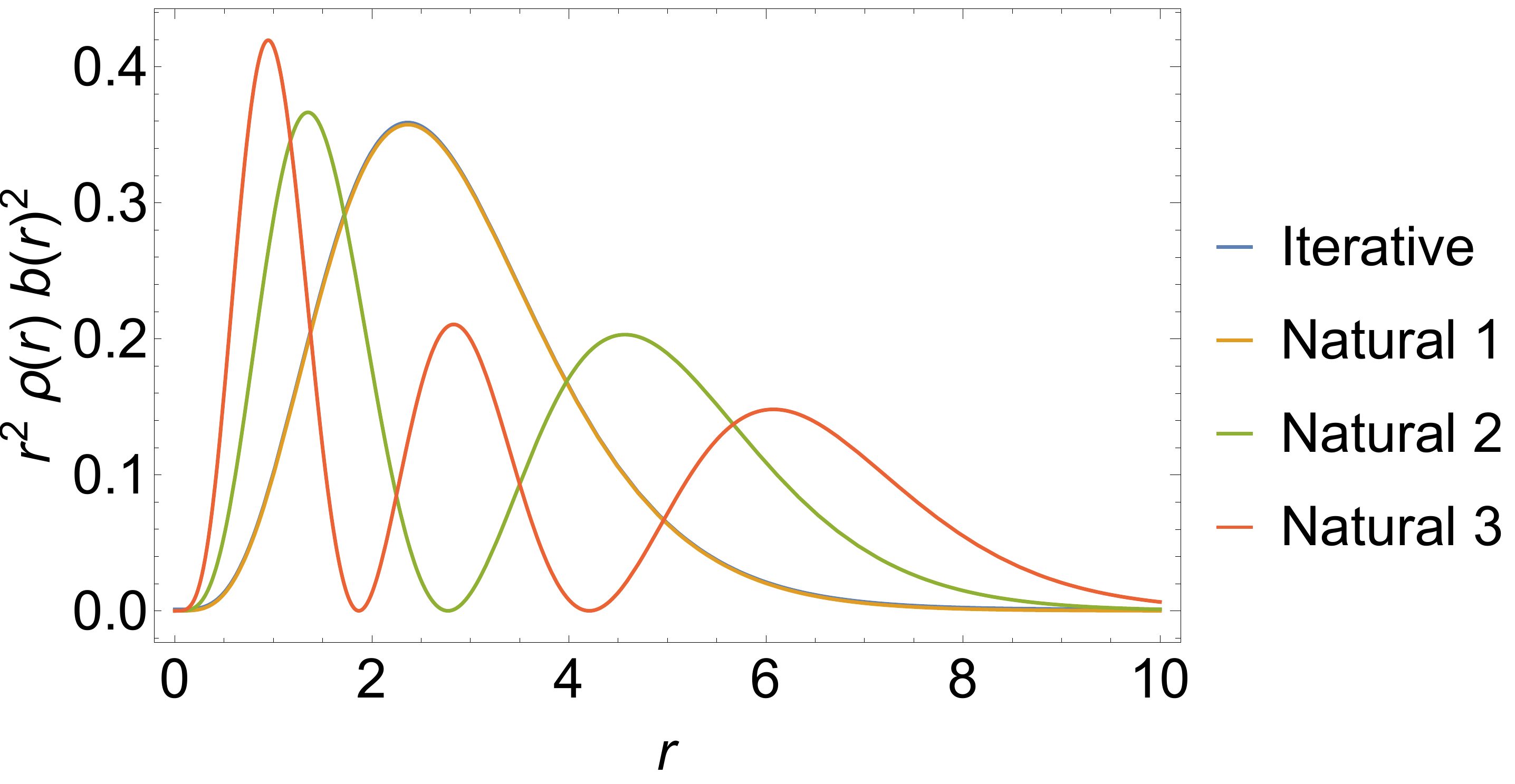}
    \includegraphics[height=5.5cm]{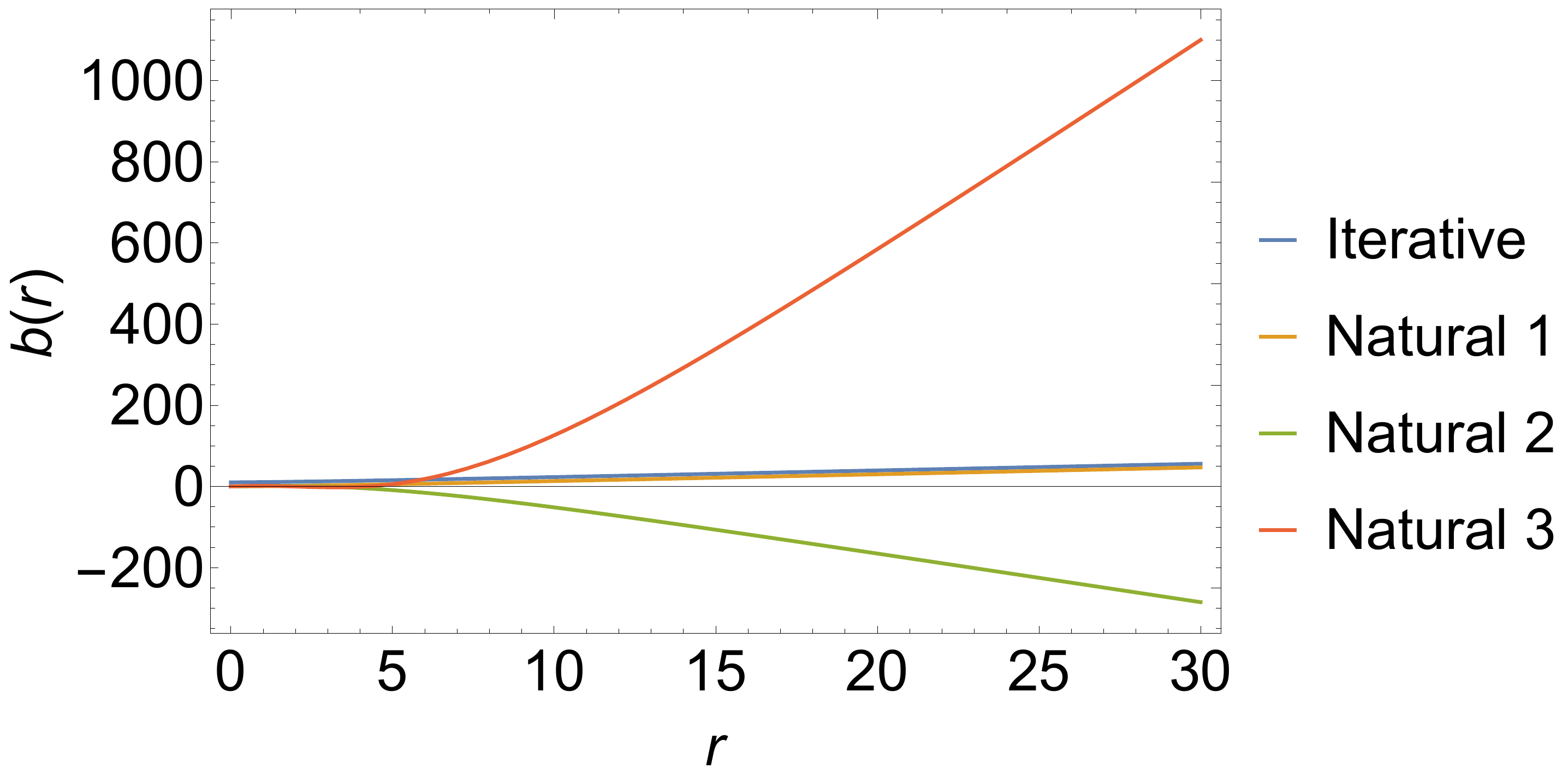}
  \caption{Comparison between the iterative solution of equation \eqref{eq:eigenvalue2} and the first three eigenfunctions of equation \eqref{eq:decomposition} for two hydrogen atoms. The iterative solution is shifted up by $0.001$ in the upper plot and $10$ in the lower plot to make it distinguishable from the first dispersals. The corresponding natural dispersal occupation numbers $n_i$ are $n_1 \approx 1.110, n_2 \approx 0.0140$ and $n_3 \approx 0.000566$.  }
  \label{fgr:bnatplot}
\end{figure}

The optimisation procedure as described is quite cumbersome and can be replaced by using instead natural dispersals $b_\mathrm{nat}$, which are obtained by performing a Singular Value Decomposition (SVD) of $J(\rv_1, \rv_2)$ into singular values $n_i$ and left- and right-singular functions $b_{\mathrm{nat}, i}^{A/B}$,
\begin{equation}
J(\rv_1, \rv_2) = \sum_{i} n_i b_{\mathrm{nat}, i}^A(\rv_1) b_{\mathrm{nat}, i}^B(\rv_2), \label{eq:decomposition}
\end{equation}
and, in the case of systems $A$ and $B$ being the same, instead one uses eigenvalue decomposition. In a basis the problem becomes that of the decomposition of the coefficient matrix $c_{ij}$, which is trivial to solve. For both $A$ and $B$ hydrogen 1s states, we show the first three natural dispersals $b_i(r)$ with highest natural dispersal occupation numbers $n_i$ in figure \ref{fgr:bnatplot}.  Interestingly, the most strongly occupied natural dispersal, matches quite closely the iterative solution and is also by itself enough to retrieve $99.97\%$ of the $C_6$ coefficient, while using the first two natural dispersals is enough for $99.999916 \%$ of $C_6$. That is, the natural dispersals provide a highly efficient way to expand $J(\rv_1, \rv_2)$. The natural dispersals also have an increasing number of nodes with decreasing occupation, analogous to natural orbitals. Finally, they all increase linearly at large $r$ as can be seen in figure \ref{fgr:bnatplot}. 

 \begin{figure}[h]
\centering
\includegraphics[height=5.5cm]{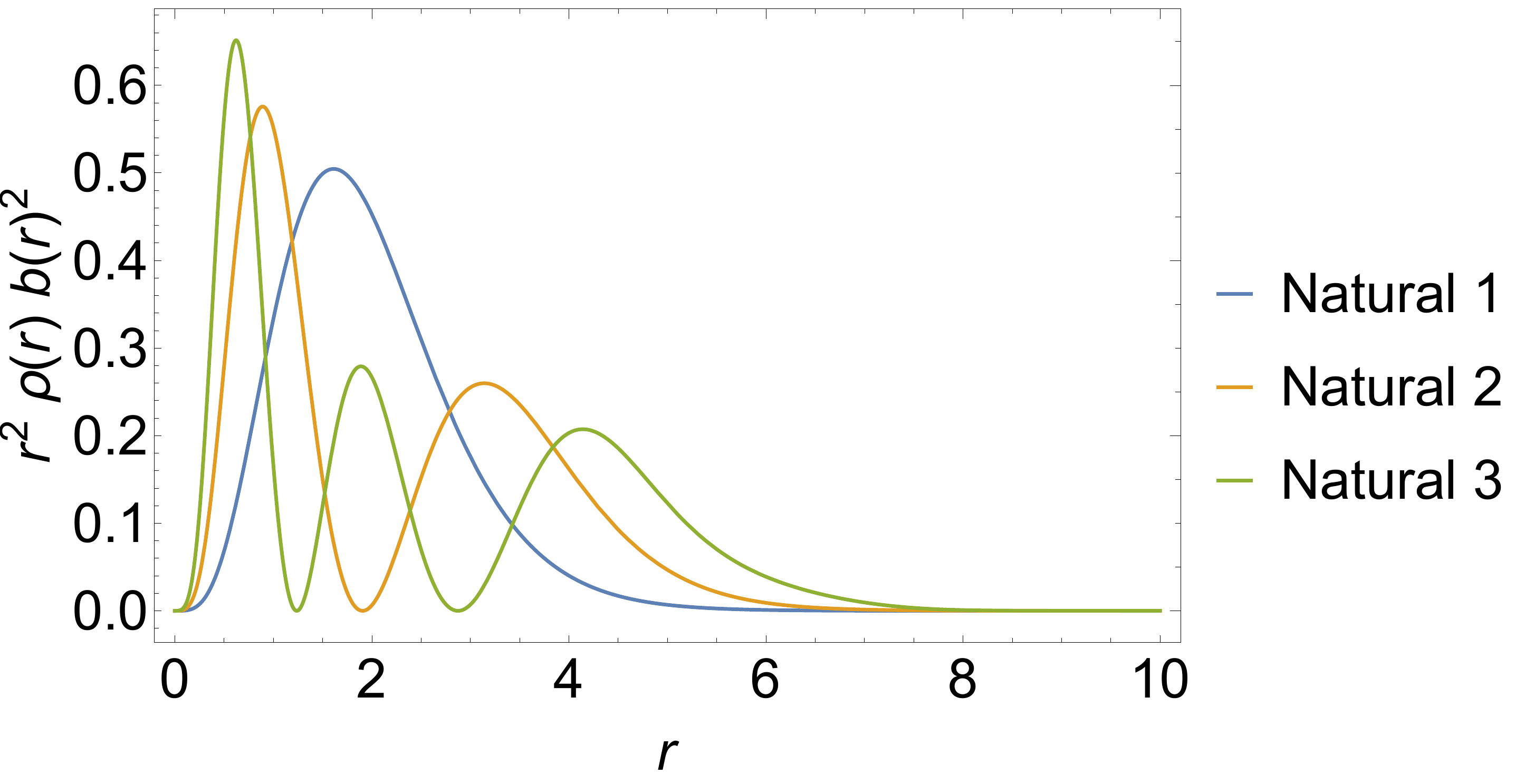}
\includegraphics[height=5.5cm]{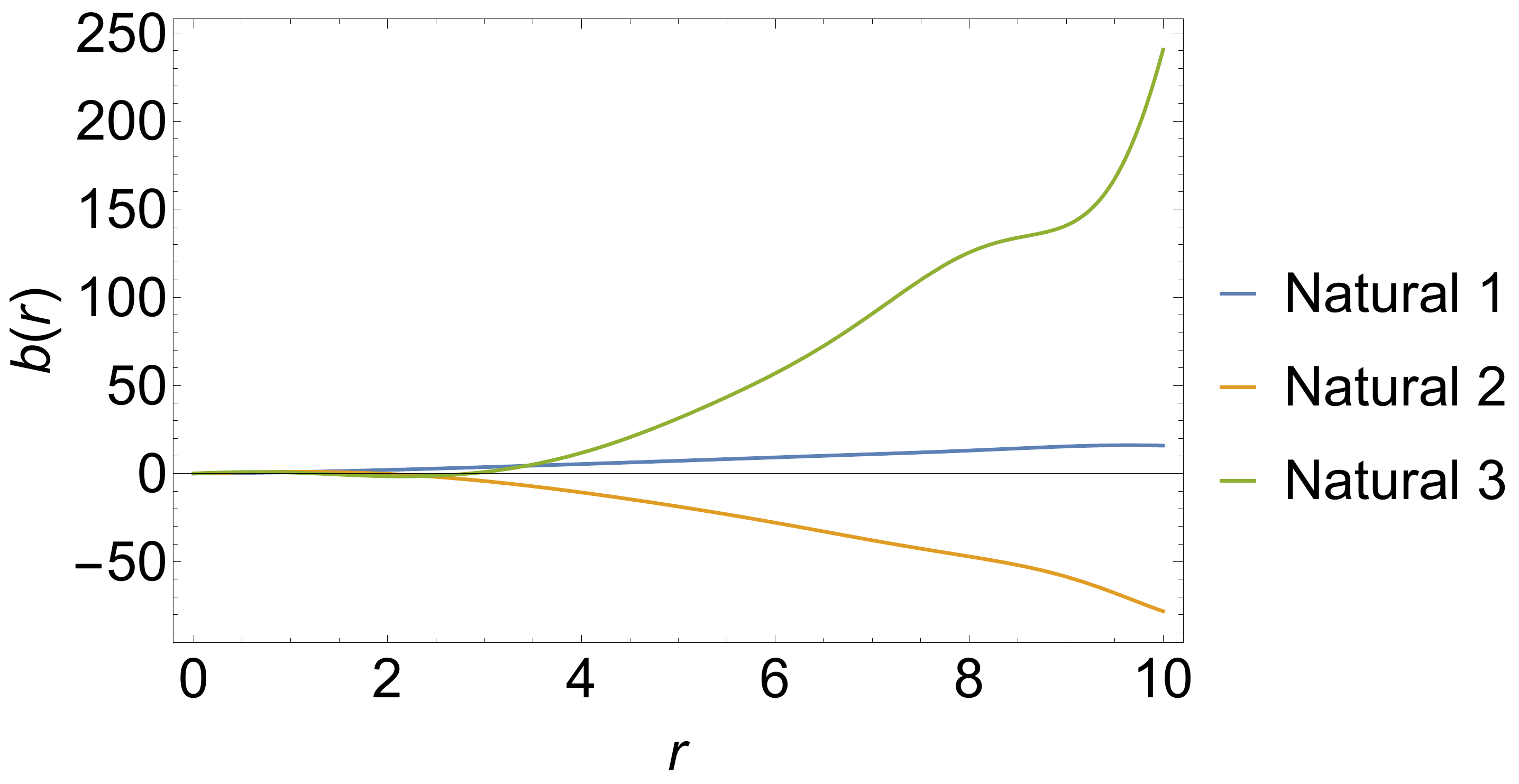}
  \caption{The first three eigenfunctions of equation \eqref{eq:decomposition} for two helium atoms. Computed from the wavefunction of Freund, et al.\cite{FreHuxMor-PRA-84}.  The corresponding natural dispersal occupation numbers $n_i$ are $n_1 \approx 0.6741, n_2 \approx 0.0148$ and $n_3 \approx 0.000782$.  }
  \label{fgr:optbHe}
\end{figure}


We show the natural dispersals for the He-He case in figure \ref{fgr:optbHe}. The calculation was performed as in  ref~\citenum{KooGor-JPCL-19}, with monomials $r^i$ for $i=0,\dots 18$. Curiously, when comparing to the natural dispersals for the hydrogen-hdrogen case of figure \ref{fgr:bnatplot} we see that they look very similar. At larger $r$ the natural dispersals for helium start to show irregular behaviour, which is a result of the density for the helium atoms not being exact.

\subsubsection{Gaussian one-electron densities (Drude model)}
A special case occurs if both $\rho^A(\rv)$ and $\rho^B(\rv)$ are spherical gaussians each containing one electron, which coincides with
the (isotropic) harmonic/Drude oscillator. In this case we obtain that the exact solution for the dipole-dipole term is given by,
\begin{equation}
J(\rv_1, \rv_2) = - \hat{H}_{\mathrm{int},3} = - x_1 x_2 - y_1 y_2 + 2 z_1 z_2,
\end{equation}
for two spherical gaussians aligned along the $z$-axis. The anisotropic case can be found by scaling the different components. 

Thus, for the spherically symmetric case with two equal gaussians, $\rho^A(\rv)=\rho^B(\rv)=(\pi/\omega)^{3/2}e^{-\omega r^2}$, one obtains, with the radial $b(r)$ defined as in eq~\eqref{eq:radialbDEF},
\begin{equation}
b(r) = \sqrt{\frac{2 \omega}{3}} r,\qquad \langle w \rangle = \frac{1}{2 \omega},\qquad \langle \tau \rangle = 2 \omega,
\end{equation} 
yielding $ C_6 =6\frac{ \langle w \rangle^2}{ \langle \tau \rangle}= \frac{3}{4 \omega^3}$,
which is the exact result for the Drude model.\cite{IpsSpl-AJP-15}  The solution $b(r) = \sqrt{\frac{2 \omega}{3}} r$ is also exactly the iterative one, obtained from equation \eqref{eq:eigenvalue} and it is also the only occupied natural dispersal. 

\section{Conclusions and Perspectives}
The idea to use a microscopic mechanism based on a simple competition between kinetic energy and fragment-fragment interaction provides an expression for the dispersion energy in terms of the monomer isolated densities and xc-holes, opening a new route to build density functional approximations. Behind these expressions there is the explicit construction of a supramolecular wavefunction constrained to leave the diagonal of the many-body density matrices of the two fragments unchanged ("fixed diagonal matrices" -- $\method$), defined in a similar way as the Levy's constrained search\cite{Lev-PNAS-79} for the DFT universal functional.\\
For closed-shell many-electron atoms and molecules, the $\method$ has been shown to provide accurate and robust $C_6$ coefficients (including anisotropies) when using CCSD xc-holes,\cite{KooWecGor-arxiv-20} showing that the reduced $\method$ variational freedom does not particularly hamper the accuracy of the dispersion energy expressions, which can be then used as a basis to build new density functional approximations.\\
The interaction energy expression in its present form is exact up to and including $\mathcal{O}(R^{-10})$ for any two spherical one-electron densities $\rho^A(\rv)$ and $\rho^B(\rv)$, regardless whether they are of the same species or not. As such, it could be also used in other frameworks, for example the one proposed by Silvestrelli,\cite{Sil-PRL-08,SilBenGruAncToi-JCP-09} where the interacting fragments are maximally localized Wannier functions (MLWF) instead of atoms. The $\method$ interaction energy should provide more accurate and well defined expressions for the dispersion interaction energy between two MLWF's or other kind of localised orbitals. \\
The challenges and possible directions ahead are several: the construction of optimal ``dispersals,'' which should probably go hand in hand with an approximation for the xc-hole, as accurate xc-hole projected dispersals is what is ultimately needed; the analysis and definition of atomic dispersion energy, by using atomic optimal dispersals as basis; finally, a self-consistent implementation, which can be made possible if good xc-hole density functionals for this framework are found.

\noindent{\it Acknowledgements --} 
Financial support from European Research Council under H2020/ERC Consolidator Grant corr-DFT (Grant Number 648932) and the Netherlands Organisation for Scientific Research under Vici grant 724.017.001 is acknowledged.

\bibliography{dispersion_bib.bib}

\end{document}